\documentclass[article,preprint,groupedaddress]{revtex4}
\usepackage{epsfig}
\usepackage{graphicx}

\begin{document}

\title{Upper bound on the radii of black-hole photonspheres}
\author{Shahar Hod}
\address{The Ruppin Academic Center, Emeq Hefer 40250, Israel}
\address{}
\address{The Hadassah Institute, Jerusalem 91010, Israel}
\date{\today}

\begin{abstract}

\ \ \ One of the most remarkable predictions of the general theory
of relativity is the existence of black-hole ``photonspheres",
compact null hypersurfaces on which massless particles can orbit the
central black hole. We prove that every spherically-symmetric
asymptotically flat black-hole spacetime is characterized by a
photonsphere whose radius is bounded from above by $r_{\gamma} \leq
3M$, where $M$ is the total ADM mass of the black-hole spacetime. It
is shown that hairy black-hole configurations conform to this upper
bound. In particular, the null circular geodesic of the (bald)
Schwarzschild black-hole spacetime saturates the bound.
\end{abstract}
\bigskip
\maketitle

\section{Introduction}

The geodesic motions of test particles around central compact
objects (e.g. black holes, neutron stars) provide valuable
information about the structure and geometry of the corresponding
curved spacetime \cite{Bar,Chan,Shap,CarC}. Of particular importance
are circular null geodesics —-- orbits with constant coordinate
radii on which massless particles can orbit the central compact
object \cite{Notebn}. Circular null orbits (also known as ``photon
orbits" or ``photonspheres") are interesting from both an
astrophysical and theoretical points of view
\cite{Bar,Chan,Shap,CarC,Hodns}. For example, the optical appearance
to external observers of a compact star undergoing gravitational
collapse is closely related to the physical properties of the
photonsphere \cite{CarC,Pod,Ame}.

Unstable circular null geodesics are also related to the
characteristic (quasinormal) resonances of black-hole spacetimes
\cite{CarC,Mash,Goeb,Hod1,Dolan,Dec,Yang}. In particular, in the
framework of the geometric-optics (eikonal) approximation, these
characteristic resonances can be interpreted in terms of massless
particles temporarily trapped at the unstable null orbit of the
black-hole spacetime \cite{CarC,Mash,Goeb,Hod1,Dolan,Dec,Yang}. In
addition, it was shown \cite{Ste} that the remarkable phenomenon of
strong gravitational lensing by black holes is closely related to
the presence of null circular geodesics in the corresponding
black-hole spacetimes.

Furthermore, for hairy black-hole spacetimes a theorem was recently
proved \cite{Hodhair,Nun} that reveals the central role played by
the black-hole photonsphere in determining the effective length of
the hair outside the black-hole horizon. According to this theorem,
the non-trivial (non-asymptotic) behavior of the hair must extend
above the black-hole photonsphere. In this respect, the black-hole
photonsphere provides a generic lower bound on the effective length
of the black-hole hair \cite{Hodhair,Nun}. In addition, it was
recently proved \cite{Hodfast} that the innermost null circular
geodesic of a black-hole spacetime provides the fastest way (as
measured by asymptotic observers) to circle the central black hole.

The central role played by photonspheres in both astrophysical
\cite{CarC,Pod,Ame} and cosmological \cite{Ste} scenarios involving
black holes, as well as in purely theoretical studies of black-hole
spacetimes
\cite{CarC,Mash,Goeb,Hod1,Dolan,Dec,Yang,Hodhair,Hodfast}, makes it
highly important to explore the physical properties of these unique
null orbits. The main goal of the present study is to extend our
knowledge about the physical properties of these fascinating
geodesics. In particular, in the present paper we shall address the
following question: In a generic black-hole spacetime, how close is
the photonsphere to the black-hole horizon? As we shall show below,
one can derive a generic {\it upper bound} on the radii of
black-hole photonspheres. This bound is expressed in terms of the
total (ADM) mass of the black-hole spacetime.

\section{Description of the system}

We consider static, spherically symmetric, asymptotically flat
black-hole spacetimes. The line element may take the following form
in Schwarzschild coordinates \cite{Nun,Hodfast,Hodm}
\begin{equation}\label{Eq1}
ds^2=-e^{-2\delta}\mu dt^2 +\mu^{-1}dr^2+r^2(d\theta^2 +\sin^2\theta
d\phi^2)\  ,
\end{equation}
where the metric functions $\delta(r)$ and $\mu(r)$ depend only on
the Schwarzschild areal coordinate $r$. Asymptotic flatness requires
\begin{equation}\label{Eq2}
\mu(r\to\infty) \to 1\ \ \ {\text{and}}\ \ \ \delta(r\to\infty) \to
0\ .
\end{equation}

It is worth emphasizing that our results would be valid for all
spherically symmetric asymptotically flat black-hole spacetimes. We
note, in particular, that we do {\it not} assume $\delta(r)=0$ (a
property which characterizes the familiar Schwarzschild and
Reissner-Nordstr\"om black-hole spacetimes) and thus our results
would be applicable to hairy black-hole configurations as well [in
these spacetimes $\delta(r)\neq 0$, see
\cite{BizCol,Lavr,Green,Stra,Volkov} and references therein].

Taking $T^{t}_{t}=-\rho$, $T^{r}_{r}=p$, and
$T^{\theta}_{\theta}=T^{\phi}_{\phi}=p_T$, where $\rho$, $p$, and
$p_T$ are identified as the energy density, radial pressure, and
tangential pressure respectively \cite{Bond1}, the Einstein
equations $G^{\mu}_{\nu}=8\pi T^{\mu}_{\nu}$ read
\begin{equation}\label{Eq3}
\mu'=-8\pi r\rho+(1-\mu)/r\  ,
\end{equation}
and
\begin{equation}\label{Eq4}
\delta'=-4\pi r(\rho +p)/\mu\  ,
\end{equation}
where the prime stands for differentiation with respect to $r$. (We
use natural units in which $G=c=1$.)

Regularity of the event horizon at $r=r_{\text{H}}$ requires
\cite{Nun}
\begin{equation}\label{Eq5}
\mu(r_{\text{H}})=0\ \ \ {\text{with}}\ \ \ \mu'(r_{\text{H}})\geq
0\ ,
\end{equation}
and
\begin{equation}\label{Eq6}
\delta(r_{\text{H}})<\infty\ \ \ ; \ \ \
\delta'(r_{\text{H}})<\infty\ .
\end{equation}
From Eqs. (\ref{Eq3}) and (\ref{Eq5}) one finds the boundary
condition \cite{Bekreg,Nun}
\begin{equation}\label{Eq7}
8\pi r^2_{\text{H}}\rho(r_{\text{H}})\leq 1.
\end{equation}
The equality sign corresponds to extremal black holes. From Eqs.
(\ref{Eq4})-(\ref{Eq6}) one finds the additional boundary condition
\begin{equation}\label{Eq8}
-p(r_{\text{H}})=\rho(r_{\text{H}})\  .
\end{equation}

The mass $m(r)$ contained within a sphere of radius $r$ is given by
\begin{equation}\label{Eq9}
m(r)={1\over2}r_{\text{H}}+\int_{r_{\text{H}}}^{r} 4\pi r'^{2}
\rho(r')dr'\ ,
\end{equation}
where $m(r_{\text{H}})=r_{\text{H}}/2$ is the horizon mass. From
Eqs. (\ref{Eq3}) and (\ref{Eq9}) one finds the relation
\begin{equation}\label{Eq10}
\mu(r)\equiv 1-2m(r)/r\  .
\end{equation}
A finite mass configuration is characterized by a density profile
$\rho(r)$ which approaches zero faster than $r^{-3}$,
\begin{equation}\label{Eq11}
r^3\rho(r)\to 0\ \ \ \text{as} \ \ \ r\to\infty\  .
\end{equation}

\section{Upper bound on the radii of black-hole photonspheres}

We shall now consider the following question: In the generic
spherically-symmetric black-hole spacetime (\ref{Eq1}), how close is
the photonsphere to the black-hole horizon? To answer this question
in the most general form, we shall now prove the existence of a
generic {\it upper bound} on the radius of the innermost null
circular geodesic.

We shall first follow the analysis of
\cite{Chan,CarC,Hodhair,Hodfast} in order to calculate the location
$r=r_{\gamma}$ of the null circular geodesic for a black-hole
spacetime described by the line element (\ref{Eq1}). The Lagrangian
describing the geodesics in the spacetime (\ref{Eq1}) is given by
\begin{equation}\label{Eq12}
2{\cal L}=-e^{-2\delta}\mu\dot t^2+\mu^{-1}\dot r^2+r^2\dot\phi^2\ ,
\end{equation}
where a dot stands for differentiation with respect to proper time.

The generalized momenta derived from the Lagrangian (\ref{Eq12}) are
given by \cite{Chan,CarC}
\begin{equation}\label{Eq13}
p_t=-e^{-2\delta}\mu\dot t\equiv -E={\text{const}}\  ,
\end{equation}
\begin{equation}\label{Eq14}
p_{\phi}=r^2\dot\phi\equiv L={\text{const}}\  ,
\end{equation}
and
\begin{equation}\label{Eq15}
p_r=\mu^{-1}\dot r\  .
\end{equation}
The Lagrangian is independent of both $t$ and $\phi$. This implies
that $E$ and $L$ are constants of the motion. The Hamiltonian of the
system is given by \cite{Chan,CarC} ${\cal H}=p_t\dot t +p_r\dot r
+p_{\phi}\dot\phi-{\cal L}$, which implies
\begin{eqnarray}\label{Eq16}
2{\cal H}=-E\dot t+L\dot\phi+\mu^{-1}\dot
r^2=\epsilon={\text{const}}\ ,
\end{eqnarray}
where $\epsilon=0$ for null geodesics and $\epsilon=1$ for timelike
geodesics. Substituting Eqs. (\ref{Eq13})-(\ref{Eq14}) into
(\ref{Eq16}), one finds
\begin{equation}\label{Eq17}
\dot
r^2=\mu\Big({{E^2}\over{e^{-2\delta}\mu}}-{{L^2}\over{r^2}}\Big)\
\end{equation}
for null geodesics.

Circular geodesics are characterized by $\dot r^2=(\dot r^2)^{'}=0$
\cite{Chan,CarC}. This yields the relation
\begin{equation}\label{Eq18}
2e^{-2\delta}\mu-r(e^{-2\delta}\mu)^{'}=0\
\end{equation}
for the null circular geodesic. Substituting the Einstein equations
(\ref{Eq3})-(\ref{Eq4}) into Eq. (\ref{Eq18}), one finds the
characteristic relation
\begin{equation}\label{Eq19}
{\cal N}(r=r_{\gamma})=0\
\end{equation}
for null circular orbits, where
\begin{equation}\label{Eq20}
{\cal N}(r)\equiv 3\mu-1-8\pi r^2p\  .
\end{equation}

We shall henceforth consider the innermost circular null geodesic of
the black-hole spacetime. Of course, this geodesic must satisfy the
characteristic equation (\ref{Eq19}). We shall first prove that such
circular null geodesic must exist in the black-hole spacetime:
taking cognizance of Eqs. (\ref{Eq5}), (\ref{Eq7}), (\ref{Eq8}), and
(\ref{Eq20}), one finds
\begin{equation}\label{Eq21}
{\cal N}(r_{\text{H}})\leq 0\
\end{equation}
at the black-hole horizon \cite{Noteex}. In addition, from Eqs.
(\ref{Eq2}), (\ref{Eq11}), and (\ref{Eq20}) [see also Eq.
(\ref{Eq28}) below] one finds the asymptotic behavior
\begin{equation}\label{Eq22}
{\cal N}(r\to\infty)\to 2\  .
\end{equation}
Inspection of Eqs. (\ref{Eq21}) and (\ref{Eq22}) reveals that there
must be at least one intermediate radius $r_{\gamma}$ (located
between the black-hole horizon and spatial infinity) for which
${\cal N}(r=r_{\gamma})=0$. This radius corresponds to the location
of the null circular geodesic, see Eq. (\ref{Eq19}).

It is worth emphasizing that the null circular geodesic (\ref{Eq19})
is the limiting case of timelike circular geodesics. That is, the
null circular geodesic is the innermost circular orbit in the
black-hole spacetime \cite{Chan,CarC}. The spacetime region between
the black-hole horizon and the photonsphere, $r_{\text{H}}\leq
r<r_{\gamma}$, in which circular geodesics are excluded, is
characterized by the inequality
\begin{equation}\label{Eq23}
{\cal N}(r_{\text{H}}\leq r<r_{\gamma})<0\  .
\end{equation}

We shall now derive a generic upper bound on the radii of the
innermost null circular geodesics. The conservation equation
$T^{\mu}_{\nu ;\mu}=0$ for the energy-momentum tensor has only one
nontrivial component \cite{Nun}
\begin{equation}\label{Eq24}
T^{\mu}_{r ;\mu}=0\  .
\end{equation}
Substituting Eqs. (\ref{Eq3}) and (\ref{Eq4}) into Eq. (\ref{Eq24}),
one finds for the pressure gradient
\begin{eqnarray}\label{Eq25}
p'(r)= {{1} \over {2\mu r}}\big[(3\mu-1-8\pi r^2p)(\rho+p)+2\mu T
-8\mu p\big]\ ,
\end{eqnarray}
where $T=-\rho+p+2p_T$ is the trace of the energy momentum tensor.
Below we shall analyze the behavior of the pressure function $P(r)
\equiv r^{4}p(r)$, whose derivative is given by
\begin{eqnarray}\label{Eq26}
P'(r)&=& {{r^{3}} \over {2\mu}}\big[{\cal N}(\rho+p)+2\mu T\big]\ .
\end{eqnarray}

When analyzing the coupled Einstein-matter system, one usually
imposes some energy conditions on the matter fields. We shall assume
that the hair (the matter fields) outside the black-hole horizon
satisfies the following conditions:
\newline
(1) The weak energy condition (WEC). This means that the energy
density is positive semidefinite,
\begin{equation}\label{Eq27}
\rho\geq 0\  ,
\end{equation}
and that it bounds the pressures. In particular, $|p| \leq \rho$.
This implies the inequality
\begin{equation}\label{Eq28}
\rho+p \geq 0\  .
\end{equation}
\newline
(2) The trace of the energy-momentum tensor plays a central role in
determining the spacetime geometry of compact objects \cite{Bond1}.
It is usually assumed to satisfy the relation $p+2p_T \leq \rho$
(see \cite{Bond1} and references therein), which implies
\begin{equation}\label{Eq29}
T\leq 0\  .
\end{equation}
\newline
It is worth emphasizing that the two conditions (\ref{Eq28}) and
(\ref{Eq29}) are indeed satisfied in all Einstein-matter models in
which hairy black-hole configurations have been found (see
\cite{Nun} for details).

From Eqs. (\ref{Eq8}) and (\ref{Eq27}) one finds the boundary
condition
\begin{equation}\label{Eq30}
P(r_{\text{H}})\leq 0\  .
\end{equation}
Furthermore, taking cognizance of the pressure gradient
(\ref{Eq26}), together with the energy conditions (\ref{Eq28}) and
(\ref{Eq29}), and the characteristic inequality (\ref{Eq23}), one
finds
\begin{equation}\label{Eq31}
P'(r_{\text{H}}\leq r<r_{\gamma})<0\
\end{equation}
below the photonsphere.

Combining the two inequalities (\ref{Eq30}) and (\ref{Eq31}), one
concludes that $P(r)$ is a non-positive and decreasing function in
the spacetime region below the photonsphere. In particular,
\begin{equation}\label{Eq32}
p(r_{\gamma})\leq 0\
\end{equation}
at the null circular geodesic. Substituting (\ref{Eq32}) into the
characteristic equations (\ref{Eq19})-(\ref{Eq20}) for the null
circular geodesic, one finds
\begin{equation}\label{Eq33}
\mu(r_{\gamma})\leq {1\over 3}\  ,
\end{equation}
which implies [see Eq. (\ref{Eq10})]
\begin{equation}\label{Eq34}
r_{\gamma}\leq 3m(r_{\gamma})\  .
\end{equation}
Finally, from Eqs. (\ref{Eq9}), (\ref{Eq27}), and (\ref{Eq34}) one
obtains the upper bound
\begin{equation}\label{Eq35}
r_{\gamma}\leq 3M\  ,
\end{equation}
where $M\equiv m(r\to\infty)$ is the total (ADM) mass of the
black-hole spacetime.

\section{Summary and discussion}

The characteristic photonspheres of spherically-symmetric black-hole
spacetimes were studied analytically within the framework of the
general theory of relativity. In particular, we have proved the
existence of a generic upper bound on the radii of the innermost
black-hole null circular geodesics [see Eq. (\ref{Eq35})]. This
bound is expressed in terms of the total ADM mass of the black-hole
spacetime. It was shown that hairy black-hole configurations conform
to this upper bound. Remarkably, the generic bound (\ref{Eq35}) is
saturated by the null circular geodesic of the (bald) Schwarzschild
black-hole spacetime.

Finally, it is worth mentioning that black-hole photonspheres are
closely related to the characteristic quasinormal resonances of the
corresponding black-hole spacetimes
\cite{CarC,Mash,Goeb,Hod1,Dolan,Dec,Yang}. In particular, in the
eikonal (geometric-optics) limit, the real parts of the resonances
are given by the simple relation \cite{CarC}
\begin{equation}\label{Eq36}
\Re\omega_m=m\times\Omega_{\infty}\  .
\end{equation}
Here $\Omega_{\infty}$ is the angular velocity (as measured by
asymptotic observers) associated with the null circular geodesic of
the black-hole spacetime, and $m\gg1$ is the azimuthal quantum
number of the perturbation mode. The angular velocity as measured by
asymptotic observers, $\Omega_{\infty}$, is given by \cite{Hodfast}
\begin{equation}\label{Eq37}
\Omega_{\infty}={{[{e^{-2\delta(r_{\gamma})}\mu(r_{\gamma})}]^{1/2}}\over{r_{\gamma}}}\
.
\end{equation}
The nominator of Eq. (\ref{Eq37}) represents the red-shift factor
$g^{1/2}_{tt}$ due to the presence of the central black hole [see
Eq. (\ref{Eq1})]. Using the inequality $\mu(r_{\gamma})\leq {1/3}$
which characterizes the black-hole photonsphere [see Eq.
(\ref{Eq33})], together with the inequalities $\delta\geq 0$
\cite{Notedel} and $r_{\gamma}\geq r_{\text{H}}$, one finds from
(\ref{Eq37}) the upper bound
\begin{equation}\label{Eq38}
\Omega_{\infty}r_{\text{H}}\leq {{1}\over{\sqrt{3}}}\
\end{equation}
on the dimensionless angular velocity associated with the null
circular geodesic of the black-hole spacetime. This bound, together
with the relation (\ref{Eq36}) between the real parts of the
black-hole resonances and the angular velocity at the null circular
geodesic, yield the interesting upper bound
\begin{equation}\label{Eq39}
\Re\omega_m r_{\text{H}}\leq m\times{{1}\over{\sqrt{3}}}\
\end{equation}
on the black-hole resonances in the eikonal $m\gg1$ limit. This
newly derived bound should be valid for generic
spherically-symmetric asymptotically flat black-hole spacetimes.

\bigskip
\noindent {\bf ACKNOWLEDGMENTS}

This research is supported by the Carmel Science Foundation. I thank
Yael Oren, Arbel M. Ongo and Ayelet B. Lata for stimulating
discussions.


\begin{thebibliography}{99}
\bibitem{Noteex} The equality sign corresponds to extremal black holes, in which
case a null circular geodesic may exist with
$r_{\gamma}=r_{\text{H}}$. We shall henceforth consider non-extremal
black holes for which ${\cal N}(r_{\text{H}})<0$.
\bibitem{Notedel} Here we have used the fact that $
\delta'\leq 0$ [see Eqs. (\ref{Eq4}) and (\ref{Eq28})] together with
the asymptotic behavior (\ref{Eq2}).
\end{thebibliography}
\end{document}